# Multipole surface solitons in layered thermal media

Yaroslav V. Kartashov, Victor A. Vysloukh*, and Lluis Torner

*ICFO-Institut de Ciencies Fotoniques, and Universitat Politecnica de Catalunya, Mediterranean Technology Park, 08860 Castelldefels (Barcelona), Spain*

We address the existence and properties of multipole solitons localized at a thermally insulating interface between uniform or layered thermal media and a linear dielectric. We find that in the case of uniform media, only surface multipoles with less than three poles can be stable. In contrast, we reveal that periodic alternation of the thermo-optic coefficient in layered thermal media makes possible the stabilization of higher order multipoles.

OCIS codes: 190.0190, 190.6135

Under appropriate conditions the nonlinear response of suitable materials can be strongly nonlocal, when material regions located far away from the laser beam waist are affected by the beam and contribute to the refractive index [1]. This may occur in materials exhibiting long-range electrostatic interactions, enhanced diffusion, transport processes, etc. For example, nonlocality occurs in liquid crystals, where it affects the interactions between beams [2,3] and results in formation of multipole solitons composed from several out-of-phase spots that can be stable only when the number of spots does not exceed four [4,5]. Thermal media may show even stronger nonlocalities [6-8] because the refractive index shape induced by light extends up to the sample borders, and it is always affected by the boundary conditions [9,10]. These conditions therefore strongly affect propagation of radiation, especially near the material surface. When the conditions for stationary near-surface propagation are met, surface waves form [11,12]. In nonlocal media such waves were studied at the interfaces of diffusive Kerr-type materials [13-15]. Recently fundamental surface waves were observed in uniform focusing thermal media near its thermally insulating boundary attracting light [16]. Such two-dimensional interfaces can also support vortices and dipoles with nodal lines perpendicular to the interface [17]. Surface waves form with defocusing thermal nonlinearities too [18,19]. The properties of one-dimensional multipole surface solitons at thermally insulating interfaces remain unexplored.

In this Letter we reveal that such states possess specific restriction on the number of poles in stable soliton: Only solitons with less than three poles can be stable. Importantly, we also found that if the thermal medium is made of layers with opposite sings of the thermo-optic coefficient [20], it can support stable surface multipoles of high orders. Light beams propagating at interface of such layered medium induce a strong nonlinear lattice that causes their dramatic reshaping with increase of power in contrast to multipoles at interfaces of uniform media.

To describe the propagation of light beams near the thermally-insulating interface of thermal medium we use a nonlinear Schrödinger equation for the dimensionless amplitude $q$ of the light field coupled to the equation for normalized temperature variation $T$:

$$i\frac{\partial q}{\partial \xi} = -\frac{1}{2}\frac{\partial^2 q}{\partial \eta^2} - \sigma(\eta)Tq, \quad \frac{\partial^2 T}{\partial \eta^2} = -|q|^2 \quad \text{for} -L \leq \eta \leq 0,$$
$$i\frac{\partial q}{\partial \xi} = -\frac{1}{2}\frac{\partial^2 q}{\partial \eta^2} \quad \text{for } \eta < -L \text{ and } \eta > 0. \tag{1}$$

Here $\eta$ is the transverse coordinate; $L$ is the width of the thermal medium; the parameter $\sigma(\eta) = \sigma_\text{b} - \sigma_\text{a}\,\text{sgn}[\sin(\pi\eta/d)]$ accounts for the variations of the thermo-optic coefficient between layers of width $d$; $\sigma_\text{a}, \sigma_\text{b}$ are the amplitude of variation and constant part of the normalized thermo-optic coefficient; $\sigma(\eta) > 0$ corresponds to focusing, while $\sigma(\eta) < 0$ corresponds to defocusing nonlinearity. We assume that the refractive index of the surrounding linear medium is equal to the unperturbed refractive index of the thermal material. Thus we solve Eq. (1) with the boundary conditions $T|_{\eta=-L} = 0$ and $\partial T/\partial \eta|_{\eta=0} = 0$, i.e. $\eta = 0$ boundary is assumed thermally insulating (for example, due to large difference of thermal conductivities of contacting materials), while $\eta = -L$ boundary is thermally stabilized by means of an external heat sink. We set $L = 25.6$ and $d = 0.5$ that is typical for experiments.

The soliton solutions (including perturbed ones) of Eq. (1) can be found in the form $q = (w + u + iv)\exp(ib\xi)$, where $w(\eta)$ is a stationary soliton solution, $u(\eta, \xi)$ and $v(\eta, \xi)$ are small perturbations that can grow with complex rate $\delta = \delta_r + i\delta_i$ upon propagation, and $b$ is the propagation constant. Substitution of the light field in such form into Eq. (1) gives in leading order equation for the stationary profile $w$, while in the next order one gets the linear eigenvalue problem

$$\delta u = -\frac{1}{2}\frac{d^2 v}{d\eta^2} - \sigma Tv + bv,$$
$$\delta v = \frac{1}{2}\frac{d^2 u}{d\eta^2} + \sigma Tu + \sigma \Delta Tw - bu, \qquad (2)$$

which holds for $-L \leq \eta \leq 0$, while outside this region, for $\eta < -L$ and $\eta > 0$, one has $\sigma = 0$. The temperature perturbation stands for $\Delta T = -2\int_{-L}^{0} G(\eta,\lambda)w(\lambda)u(\lambda)d\lambda$, with the response function $G(\eta,\lambda) = -(\eta+L)$ for $\eta \leq \lambda$ and $G(\eta,\lambda) = -(\lambda+L)$ for $\eta \geq \lambda$.

First, we consider the properties of multipoles at the edge of uniform thermal media when $\sigma_a = 0$, $\sigma_b = 1$. Representative shapes of such solitons are shown in Fig. 1. Such solitons reside mostly inside the thermal medium and only weakly penetrate into the linear region. The refractive index profile $\sigma T$ depicted in Fig. 1(d) explains the formation of a thermal lens that results in concentration of light in the vicinity of interface. Note that the refractive index is nonzero in the entire thermal medium and decreases almost linearly toward its left border. In all multipoles in uniform media, the pole located farthest from the interface always exhibits a higher amplitude. The higher the number of poles the larger the power $U = \int_{-\infty}^{\infty} |q|^2 \, d\eta$ carried by solitons at a fixed $b$. For lowest-order surface states, the power increases almost linearly with $b$ [Fig. 2(a)]. The refractive index distribution broadens with increase of number of poles $n$. Increasing $b$ results in the simultaneous contraction of all poles toward the interface and in the increase in amplitude and peak refractive index at the interface [Figs. 1(a)-1(c)].

One of the central results of this Letter is that only multipoles with number of poles $n \leq 2$ were found to be stable in the case of thermally insulating interface in uniform media. This results resembles the constraint $n \leq 4$ on the number of poles in stable multipoles in natural nonlocal materials found in [5]. Notice, however, that the presence of the interface results in a more dramatic restriction on a number of spots that can be packed together into stable states. In the case of surface solitons with $n > 2$ the perturbation growth rate increases monotonically with $b$, at least far from the cutoff where $\delta_r(b)$ may exhibit a non-monotonic behavior [Fig. 2(b)]. Also, $\delta_r$ increases with $n$ at a fixed $b$.

Importantly, a periodic variation of thermo-optic coefficient was found to dramatically modify the properties of surface solitons. Here we consider the case of alternating focusing and defocusing layers, i.e. $\sigma_a = 1$, $\sigma_b = 0$. A light beam entering such a medium self-induces a nonlinear lattice that becomes more and more pronounced with increasing peak amplitude, and that is strongly asymmetric because of the boundary conditions [Fig. 3(d)]. In that case the soliton peak may be localized in any of the focusing layers. Fundamental surface solitons residing in the first layer exhibit pronounced oscillations on their left wings [Fig. 3(a)]. Increasing the peak amplitude results in concentration of light almost within a single surface layer. Multipoles with $n > 1$ centered at intermediate and high powers in second, third, etc focusing layers carry the number of poles equal to the number of layer where the most pronounced peak is located [Figs. 3(b) and 3(c)]. The pole amplitudes decrease toward the interface much faster than in uniform medium. All multipoles feature strong in-phase oscillations on their left wings. The power is a monotonically increasing function of $b$ [Fig. 4(a)]. When power decreases, the left outermost pole shifts into the bulk of the thermal medium, gradually jumping between neighboring focusing layers. This is readily visible in Fig. 3(c). In contrast, when power increases light tends to concentrate almost within a single layer with number $n$, i.e. out-of-phase oscillations in layers $n-1$, $n-2$,... also gradually vanish and one gets transformation of multipole soliton into fundamental-like looking state with only small wings. Interestingly, the envelope of the refractive index $\sigma T$ for such solitons exhibit almost flat plateau between the layer where soliton is located and interface.

We found that the periodic modulation of the thermo-optic coefficient dramatically modifies stability of the multipole surface solitons. In such periodic media multipoles with higher number of spots were found to be stable, provided that the light power exceeds a threshold value. Thus, tripole solitons that were unstable in uniform thermal medium, in layered medium exhibit only a single narrow instability domain [Fig. 4(b)]. This is the case for all multipoles with $n$ up to $10$ that we considered here. Notice that the complexity of the structure of instability domain, which for solitons with large number of spots is usually located close to low-power cutoff, increases with larger number of poles [see Fig. 4(c) showing instability domains for soliton with $n = 10$]. The results of such linear stability analysis were always confirmed by direct simulations of propagation of perturbed multipoles in Eq. (1). Figure 5 shows examples of evolution of unstable [Fig. 5(a)] and stable [Fig. 5(b)] tripole solitons. Development of instability causes pronounced amplitude oscillations and transformation of the internal structure of unstable soliton, but since thermally insulating interface tends to attract light, there is almost no radiation emission. Perturbed stable multipole propagate without any noticeable shape deformations.

Summarizing, we showed that addition of periodic modulation of thermo-optic coefficient substantially modifies the properties and shapes of surface states localized near thermally insulating boundary of composite thermal medium. While a thermally insulating edge of uniform thermal medium can not support stable solitons with more than two poles, we found that one can made stable higher-order multipoles in layered thermal media.

*Visiting from Universidad de las Americas – Puebla, Mexico.

# References with titles

1. W. Krolikowski, O. Bang, N. I. Nikolov, D. Neshev, J. Wyller, J. J. Rasmussen, and D. Edmundson, "Modulational instability, solitons and beam propagation in spatially nonlocal nonlinear media," J. Opt. B: Quantum and Semiclass. Opt. **6**, S288 (2004).
2. M. Peccianti, C. Conti, and G. Assanto, "Interplay between nonlocality and nonlinearity in nematic liquid crystals," Opt. Lett. **30**, 415 (2005).
3. M. Peccianti, K. A. Brzdakiewicz, and G. Assanto, "Nonlocal spatial soliton interactions in nematic liquid crystals," Opt. Lett. **27**, 1460 (2002).
4. X. Hutsebaut, C. Cambournac, M. Haelterman, A. Adamski, and K. Neyts, "Single-component higher-order mode solitons in liquid crystals," Opt. Comm. **233**, 211 (2004).
5. Z. Xu, Y. V. Kartashov, and L. Torner, "Upper threshold for stability of multipole-mode solitons in nonlocal nonlinear media," Opt. Lett. **30**, 3171 (2005).
6. C. Rotschild, O. Cohen, O. Manela, M. Segev, and T. Carmon, "Solitons in nonlinear media with an infinite range of nonlocality: first observation of coherent elliptic solitons and of vortex-ring solitons," Phys. Rev. Lett. **95**, 213904 (2005).
7. C. Rotschild, M. Segev, Z. Xu, Y. V. Kartashov, L. Torner, and O. Cohen, "Two-dimensional multipole solitons in nonlocal nonlinear media," Opt. Lett. **31**, 3312 (2006).
8. C. Rotschild, B. Alfassi, O. Cohen, and M. Segev, "Long-range interactions between optical solitons," Nat. Phys. **2**, 769 (2006).
9. B. Alfassi, C. Rotschild, O. Manela, M. Segev, and D. N. Christodoulides, "Boundary force effects exerted on solitons in highly nonlocal nonlinear media," Opt. Lett. **32**, 154 (2007).
10. A. Minovich, D. Neshev, A. Dreischuh, W. Krolikowski, and Y. Kivshar, "Experimental reconstruction of nonlocal response of thermal nonlinear optical media," Opt. Lett. **32**, 1599 (2007).
11. Nonlinear surface electromagnetic phenomena, Ed. by H. E. Ponath and G. I. Stegeman, North Holland, Amsterdam (1991).

12. D. Mihalache, M. Bertolotti, and C. Sibilia, "Nonlinear wave propagation in planar structures," Progr. Opt. **27**, 229 (1989).
13. P. Varatharajah, A. Aceves, J. V. Moloney, D. R. Heatley, and E. M. Wright, "Stationary nonlinear surface waves and their stability in diffusive Kerr media," Opt. Lett. **13**, 690 (1988).
14. D. R. Andersen, "Surface-wave excitation at the interface between diffusive Kerr-like nonlinear and linear media," Phys. Rev. A **37**, 189 (1988).
15. Y. V. Kartashov, L. Torner, and V. A. Vysloukh, "Lattice-supported surface solitons in nonlocal nonlinear media," Opt. Lett. **31**, 2595 (2006).
16. B. Alfassi, C. Rotschild, O. Manela, M. Segev, and D. N. Christodoulides, "Nonlocal surface-wave solitons," Phys. Rev. Lett. **98**, 213901 (2007).
17. F. Ye, Y. V. Kartashov, and L. Torner, "Nonlocal surface dipoles and vortices," Phys. Rev. A **77**, 033829 (2008).
18. Y. V. Kartashov, F. Ye, V. A. Vysloukh, and L. Torner, "Surface waves in defocusing thermal media," Opt. Lett. **32**, 2260 (2007).
19. Y. V. Kartashov, V. A. Vysloukh, and L. Torner, "Ring surface waves in thermal nonlinear media," Opt. Express **15**, 16216 (2007).
20. Y. V. Kartashov, V. A. Vysloukh, and L. Torner, "Propagation of solitons in thermal media with periodic nonlinearity," Opt. Lett. **33**, 1774 (2008).

# References without titles

1. W. Krolikowski, O. Bang, N. I. Nikolov, D. Neshev, J. Wyller, J. J. Rasmussen, and D. Edmundson, J. Opt. B: Quantum and Semiclass. Opt. **6**, S288 (2004).
2. M. Peccianti, C. Conti, and G. Assanto, Opt. Lett. **30**, 415 (2005).
3. M. Peccianti, K. A. Brzdakiewicz, and G. Assanto, Opt. Lett. **27**, 1460 (2002).
4. X. Hutsebaut, C. Cambournac, M. Haelterman, A. Adamski, and K. Neyts, Opt. Comm. **233**, 211 (2004).
5. Z. Xu, Y. V. Kartashov, and L. Torner, Opt. Lett. **30**, 3171 (2005).
6. C. Rotschild, O. Cohen, O. Manela, M. Segev, and T. Carmon, Phys. Rev. Lett. **95**, 213904 (2005).
7. C. Rotschild, M. Segev, Z. Xu, Y. V. Kartashov, L. Torner, and O. Cohen, Opt. Lett. **31**, 3312 (2006).
8. C. Rotschild, B. Alfassi, O. Cohen, and M. Segev, Nat. Phys. **2**, 769 (2006).
9. B. Alfassi, C. Rotschild, O. Manela, M. Segev, and D. N. Christodoulides, Opt. Lett. **32**, 154 (2007).
10. A. Minovich, D. Neshev, A. Dreischuh, W. Krolikowski, and Y. Kivshar, Opt. Lett. **32**, 1599 (2007).
11. Nonlinear surface electromagnetic phenomena, Ed. by H. E. Ponath and G. I. Stegeman, North Holland, Amsterdam (1991).
12. D. Mihalache, M. Bertolotti, and C. Sibilia, Progr. Opt. **27**, 229 (1989).
13. P. Varatharajah, A. Aceves, J. V. Moloney, D. R. Heatley, and E. M. Wright, Opt. Lett. **13**, 690 (1988).
14. D. R. Andersen, Phys. Rev. A **37**, 189 (1988).
15. Y. V. Kartashov, L. Torner, and V. A. Vysloukh, Opt. Lett. **31**, 2595 (2006).
16. B. Alfassi, C. Rotschild, O. Manela, M. Segev, and D. N. Christodoulides, Phys. Rev. Lett. **98**, 213901 (2007).
17. F. Ye, Y. V. Kartashov, and L. Torner, Phys. Rev. A **77**, 033829 (2008).
18. Y. V. Kartashov, F. Ye, V. A. Vysloukh, and L. Torner, Opt. Lett. **32**, 2260 (2007).
19. Y. V. Kartashov, V. A. Vysloukh, and L. Torner, Opt. Express **15**, 16216 (2007).

20. Y. V. Kartashov, V. A. Vysloukh, and L. Torner, Opt. Lett. **33**, 1774 (2008).

# Figure captions

Figure 1.     Profiles of (a) fundamental, (b) dipole, (c) tripole solitons, and (d) refractive index distributions for fundamental solitons with different $b$ values. In all cases $\sigma_a = 0$, $\sigma_b = 1$. In gray regions $\sigma > 0$.

Figure 2.     (a) Energy flow versus $b$ for fundamental surface soliton. Circles correspond to profiles shown in Fig. 1(a). (b) $\delta_r$ versus $b$ for higher-order solitons with different number of poles. In all cases $\sigma_a = 0$, $\sigma_b = 1$.

Figure 3.     Profiles of (a) fundamental, (b) dipole, (c) tripole solitons with different $b$ values, and (d) refractive index distributions for fundamental solitons with $b = 5$ (black curve) and $b = 1$ (red curve). In all cases $\sigma_a = 1$, $\sigma_b = 0$. In gray regions $\sigma > 0$.

Figure 4.     (a) Energy flow versus $b$ for fundamental surface soliton. Circles correspond to profiles shown in Fig. 3(a). $\delta_r$ versus $b$ for solitons with $n = 3$ (b) and $n = 10$ (c). In all cases $\sigma_a = 1$, $\sigma_b = 0$.

Figure 5.     Propagation of perturbed tripole solitons with (a) $b = 7.4$ and (b) $b = 8.0$. In both cases $\sigma_a = 1$, $\sigma_b = 0$. White dashed line indicates interface position.

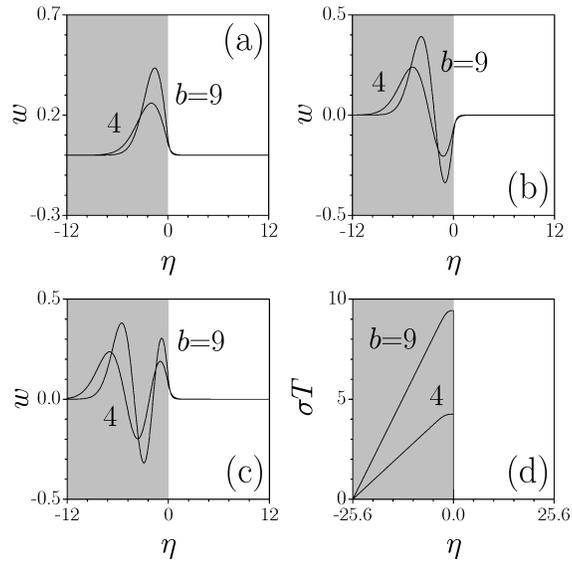

Figure 1. Profiles of (a) fundamental, (b) dipole, (c) tripole solitons, and (d) refractive index distributions for fundamental solitons with different $b$ values. In all cases $\sigma_a = 0$, $\sigma_b = 1$. In gray regions $\sigma > 0$.

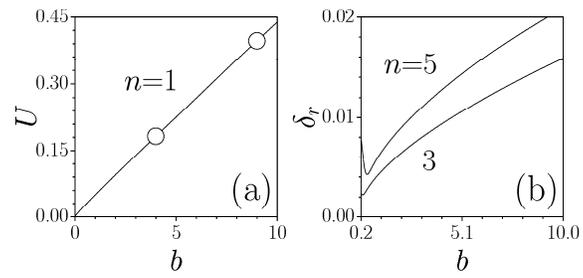

Figure 2. (a) Energy flow versus $b$ for fundamental surface soliton. Circles correspond to profiles shown in Fig. 1(a). (b) $\delta_r$ versus $b$ for higher-order solitons with different number of poles. In all cases $\sigma_\mathrm{a} = 0$, $\sigma_\mathrm{b} = 1$.

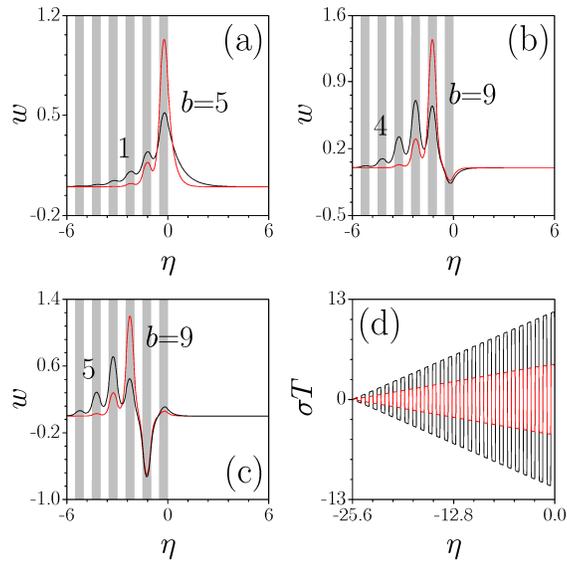

Figure 3. Profiles of (a) fundamental, (b) dipole, (c) tripole solitons with different $b$ values, and (d) refractive index distributions for fundamental solitons with $b = 5$ (black curve) and $b = 1$ (red curve). In all cases $\sigma_{\mathrm{a}} = 1$, $\sigma_{\mathrm{b}} = 0$. In gray regions $\sigma > 0$.

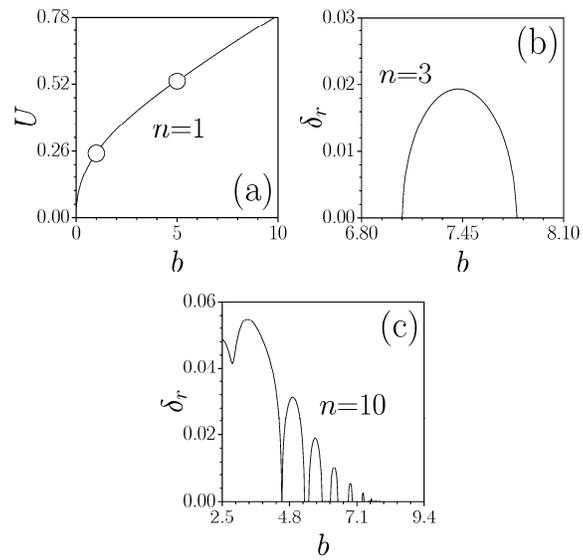

Figure 4. (a) Energy flow versus $b$ for fundamental surface soliton. Circles correspond to profiles shown in Fig. 3(a). $\delta_r$ versus $b$ for solitons with $n=3$ (b) and $n=10$ (c). In all cases $\sigma_{\rm a}=1$, $\sigma_{\rm b}=0$.

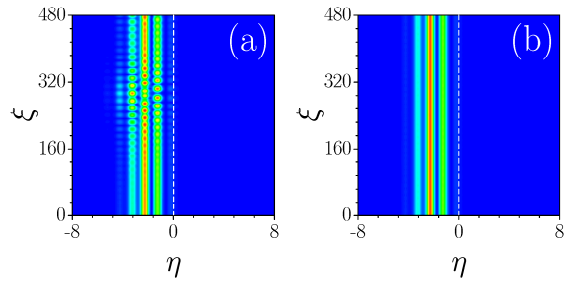

Figure 5.  Propagation of perturbed tripole solitons with (a) $b = 7.4$ and (b) $b = 8.0$. In both cases $\sigma_\mathrm{a} = 1$, $\sigma_\mathrm{b} = 0$. White dashed line indicates interface position.